\documentclass[twocolumn,showpacs,aps]{revtex4}
\usepackage{graphicx}
\usepackage{dcolumn}
\usepackage{bm}
\usepackage{amsmath}
\usepackage{amsfonts}
\usepackage{amssymb}
\usepackage{color}

\providecommand{\U}[1]{\protect\rule{.1in}{.1in}}

\begin{document}

\title{Thermodynamics of $N$-dimensional quantum walks}
\author{Alejandro Romanelli$^{(1)}$, Ra\'ul Donangelo$^{(1)}$, Renato Portugal$^{(2)}$, and Franklin de Lima Marquezino$^{(3)}$}

\affiliation{$^{(1)}$Instituto de F\'{\i}sica, Facultad de Ingenier\'{\i}a, Universidad de la Rep\'ublica, C.C. 30, C.P.
11000, Montevideo, Uruguay.}

\affiliation{$^{(2)}$Laborat\'orio Nacional de Computa\c{c}\~ao Cient\'{\i}fica, Petr\'opolis, RJ, 25651-075, Brazil.}

\affiliation{$^{(3)}$Universidade Federal do Rio de Janeiro, RJ, 21941-972, Brazil.}

\date{\today }

\begin{abstract}
The entanglement between the position and coin state of a $N$-dimensional quantum walker is shown to lead to a
thermodynamic theory. The entropy, in this thermodynamics, is associated to the reduced density operator for the
evolution of chirality, taking a partial trace over positions. From the asymptotic reduced density matrix it is possible
to define thermodynamic quantities, such as the asymptotic entanglement entropy, temperature, Helmholz free energy, etc.
We study in detail the case of a $2$-dimensional quantum walk, in the case of two different initial conditions: a
non-separable coin-position initial state, and a separable one. The resulting entanglement temperature is presented as
function of the parameters of the system and those of the initial conditions.
\end{abstract}

\maketitle

\section{Introduction}\label{intro}

The coined quantum walk (QW) model on the line was introduced by Aharonov \textit{et al.}~\cite{Aharonov} and its properties
on graphs were studied in Ref.~\cite{AAKV01}. In this model, the particle jumps from site to site in a direction which
depends on the value of an internal degree of freedom called chirality. Quantum walks on multi-dimensional lattices were
studied by many authors~\cite{MBSS02,Tregenna1,OPD06,Watabe08} and display the key feature of spreading quadratically
faster in terms of probability distribution, compared to the classical random walk model on the same underlying
structure~\cite{AF02}. Those models were successfully applied to develop quantum algorithms, specially for searching a
marked node in graphs~\cite{SKW03,AKR05,PortugalBook}. There are other models of quantum walks and some of them do not
use an auxiliary Hilbert space and have no coin. The continuous-time quantum walk model introduced by Farhi and
Gutman~\cite{FG98} and the coinless quantum walk model introduced by Patel \textit{et al.}~\cite{PRR05a} are examples of
such models. The latter model can be used to search a marked node on two-dimensional finite lattices with the same number
of steps (asymptotically in terms of the system size) compared to the coined model, with the advantage of using a smaller
Hilbert space~\cite{APN13}.

The thermodynamics of quantum walks on the line was introduced in Refs.~\cite{alejo2010,alejo2012} using the coined QW
model, which has two subspaces, namely, the coin and spatial parts. Taking the model's whole Hilbert space, the dynamics
is unitary with no change in the entropy. On the other hand, the coin subspace evolves entangled with its environment. In
the asymptotic limit ($t\rightarrow\infty$), after tracing out the spatial part, the coin reaches a final equilibrium
state which, if we consider the quantum canonical ensemble, can be seen to have an associated temperature. This procedure
allows the introduction of thermodynamical quantities and helps to understand the physics behind the dynamics. In most
cases, the thermodynamical quantities depend on the initial condition in stark contrast with the classical Markovian
behavior.

In general the Hilbert space of a quantum mechanical model factors as a tensor product $\mathcal{H}_{sys}\otimes
\mathcal{H}_{env}$ of the spaces describing the degrees of freedom of the system and environment. The evolution of the
system is determined by the reduced density operator that results from taking the trace over $\mathcal{H}_{env}$ to
obtain $\varrho_{sys}=\mathrm{tr}_{env}(\rho )$. The simple toy models similar to our model studied in Refs.
\cite{Zurek,Meyer} shows how the correlations of a quantum system with other systems may cause one of its
observables to behave in a classical manner. In this sense the fact that the partial trace over the QW positions leads to
a system effectively in thermal equilibrium, agrees with those previous results.

In this work, we focus our attention on the thermodynamics of coined quantum walks on multi-dimensional lattices. The
analysis of the dynamics is greatly simplified by using the Fourier basis (momentum space). In the computational basis,
the evolution operator is in a Hilbert space of infinite dimensions, while in the Fourier basis we use a new operator in
the finite coin subspace. The temperature of the quantum walk is obtained by taking the asymptotic limit
($t\rightarrow\infty$) of the reduced density matrix of the coin subspace and by making a correspondence to a quantum
canonical ensemble. Using the saddle point expansion theorem~\cite{BO78}, we obtain the expression of the entanglement
temperature in terms of the coin entries and the initial state. That analysis generalizes the results of
Ref.~\cite{alejo2012} and allows to obtain many new examples due to the increased number of degrees of freedom.

The paper is organized as follows. In Sec.\ref{theory} we review the dynamics of multi-dimensional coined quantum walks
in terms of the Fourier basis. In Sec.\ref{thermo} we describe the thermodynamics of quantum walks in lattices and show
how to obtain the temperature and other thermodynamical quantities. In Sec.\ref{initial} we obtain an explicit expression
for the temperature in terms of the initial condition. In Sec.\ref{examples2D} we give some examples in two dimensions.
In the last section we draw the conclusions.

\section{$N$-dimensional discrete quantum walks.}\label{theory}

In this section, following Ref. \cite{german2013}, we present a brief
theoretical development to obtain the wave function of the system.

The system moves at discrete time steps $t\in\mathbb{N}$ across an $N$%
-dimensional lattice of sites $\mathbf{x}\equiv\left(x_{1},\ldots,x_{N}%
\right)\in\mathbb{Z}^{N}$. Its evolution is governed by an unitary time
operator. This operator can be written as the application of two more simple
operators, one representing the unitary operator due to the $2N$-dimensional
coin which determines the direction of displacement and another being
specifically the unitary operator of the displacement. The Hilbert space of
the whole system has then the form
\begin{equation}
\mathcal{H}=\mathcal{H}_{\mathrm{P}}\otimes\mathcal{H}_{\mathrm{C}},
\label{espacio}
\end{equation}
where the position space, $\mathcal{H}_{\mathrm{P}}$, is spanned by the
unitary vectors $\left\{ \left\vert \mathbf{u_\alpha}\right\rangle
\equiv\left\vert
\delta_{1\alpha},\ldots,\delta_{N\alpha}\right\rangle;\alpha=1,\ldots,N%
\right\} $, and the coin space, $\mathcal{H}_{\mathrm{C}}$, is spanned by $%
2N $ orthonormal quantum states $\left\{ \left\vert
\alpha_{\eta}\right\rangle :\alpha=1,\ldots,N;\eta=\pm\right\} $. Therefore $%
\alpha$ is associated with the axis and $\eta$ with the direction. In the
usual QW on the line ($N=1$), $\left\vert 1_{-}\right\rangle $ and $%
\left\vert 1_{+}\right\rangle $ are the right and left states $\left\vert
\mathrm{R}\right\rangle $ and $\left\vert \mathrm{L}\right\rangle $. The
state of the system at any time $t$ is represented by the ket $\left\vert
\psi_{t}\right\rangle $ which can be expressed as
\begin{equation}
\left\vert \psi_{t}\right\rangle =\sum_{\mathbf{x\in\mathbb{Z}^{N}}%
}\sum_{\alpha=1}^{N}\sum_{\eta=\pm}\psi_{\mathbf{x},t}^{\alpha,\eta}\
\left\vert \mathbf{x}\right\rangle \otimes\left\vert
\alpha_{\eta}\right\rangle ,  \label{psi}
\end{equation}
where
\begin{equation}
\psi_{\mathbf{x},t}^{\alpha,\eta}=\left(\left\langle \alpha_{\eta}\right\vert \otimes\left\langle \mathbf{x}\right\vert
\right)\left\vert \psi_{t}\right\rangle .  \label{proj}
\end{equation}
We define, at each point $\mathbf{x}$, the following ket,
\begin{equation}
\left\vert \psi_{\mathbf{x},t}\right\rangle =\left\langle \mathbf{x}%
\right.\left\vert \psi_{t}\right\rangle
=\sum_{\alpha=1}^{N}\sum_{\eta=\pm}\psi_{\mathbf{x},t}^{\alpha,\eta}\left%
\vert \alpha_{\eta}\right\rangle ,  \label{psi_x}
\end{equation}
which is a coin state, so that
\begin{equation}
\psi_{\mathbf{x},t}^{\alpha,\eta}=\left\langle
\alpha_{\eta}\right.\left\vert \psi_{\mathbf{x},t}\right\rangle .
\label{psi2_x}
\end{equation}
As $\left\vert \psi_{\mathbf{x},t}^{\alpha,\eta}\right\vert ^{2}=\left\vert
\left(\left\langle \alpha_{\eta}\right\vert \otimes\left\langle \mathbf{x}%
\right\vert \right)\left\vert \psi_{t}\right\rangle \right\vert ^{2}$ is the
probability of finding the walker at $\left(\mathbf{x},t\right)$ and the
coin in state $\left\vert \alpha_{\eta}\right\rangle $, the probability of
finding the walker at $\left(\mathbf{x},t\right)$ irrespectively of the coin
state is then
\begin{equation}
P_{\mathbf{x},t}=\sum_{\alpha=1}^{N}\sum_{\eta=\pm}\left\vert \psi_{\mathbf{x%
},t}^{\alpha,\eta}\right\vert ^{2}=\left\langle \psi_{\mathbf{x}%
,t}\right.\left\vert \psi_{\mathbf{x},t}\right\rangle ,  \label{prob}
\end{equation}
where we used the fact that $\sum_{\alpha=1}^{N}\sum_{\eta=\pm}\left\vert
\alpha_{\eta}\right\rangle \left\langle \alpha_{\eta}\right\vert $ is the
identity in $\mathcal{H}_{\mathrm{C}}$. Clearly $\sum_{\mathbf{x}}P_{\mathbf{%
x},t}=1$ because $\sum_{\mathbf{x}}\left\vert \mathbf{x}\right\rangle
\left\langle \mathbf{x}\right\vert $ is the identity in $\mathcal{H}_{%
\mathrm{P}}$.

The dynamical evolution of the system is ruled by
\begin{equation}
\left\vert \psi_{t+1}\right\rangle ={\hat{U}}\left\vert
\psi_{t}\right\rangle ,  \label{map}
\end{equation}
where the unitary operator
\begin{equation}
\hat{U}=\hat{D}\circ\left(\hat{I}\otimes\hat{C}\right),  \label{U}
\end{equation}
is given in terms of the identity operator in $\mathcal{H}_{\mathrm{P}}$, $%
\hat{I}$, and two more unitary operators. First, the so-called coin operator
$\hat{C}$, which acts in $\mathcal{H}_{\mathrm{C}}$, can be written in its
more general form as
\begin{equation}
\hat{C}=\sum_{\alpha=1}^{N}\sum_{\eta=\pm}\sum_{\alpha^{\prime}=1}^{N}\sum_{%
\eta^{\prime}=\pm}C_{\alpha^{\prime},\eta^{\prime}}^{\alpha,\eta}\left\vert
\alpha_{\eta}\right\rangle \left\langle
\alpha_{\eta^{\prime}}^{\prime}\right\vert ,  \label{C}
\end{equation}
where the matrix elements $C_{\alpha^{\prime},\eta^{\prime}}^{\alpha,\eta}%
\equiv\left\langle \alpha_{\eta}\right\vert \hat{C}\left\vert
\alpha_{\eta^{\prime}}^{\prime}\right\rangle $ can be arranged as a $%
2N\times2N$ unitary square matrix $C$. Then, $\hat{D}$ is the conditional
displacement operator in $\mathcal{H}$
\begin{equation}
\hat{D}=\sum_{\mathbf{x}}\sum_{\alpha=1}^{N}\sum_{\eta=\pm}\left\vert
\mathbf{x}+\eta\mathbf{u}_{\alpha}\right\rangle \left\langle \mathbf{x}%
\right\vert \otimes\left\vert \alpha_{\eta}\right\rangle \left\langle
\alpha_{\eta}\right\vert .  \label{D}
\end{equation}
Note that, depending on the coin state $\left\vert
\alpha_{\eta}\right\rangle $, the walker moves one site to the positive or
negative direction of $x_{\alpha}$ if $\eta=+$ or $\eta=-$, respectively.

Projecting Eq.(\ref{map}) onto $\left\langle \mathbf{x}\right\vert $ and
using Eqs.(\ref{proj}),(\ref{U})--(\ref{D}) we obtain
\begin{equation}
\left\vert \psi_{\mathbf{x},t+1}\right\rangle
=\sum_{\alpha=1}^{N}\sum_{\eta=\pm}\left\vert \alpha_{\eta}\right\rangle
\left\langle \alpha_{\eta}\right\vert \hat{C}\left\vert \psi_{\mathbf{x}-\eta%
\mathbf{u}_{\alpha},t}\right\rangle ,  \label{mapket_x}
\end{equation}
which further projected onto $\left\langle \alpha_{\eta}\right\vert $ leads
to
\begin{equation}
\psi_{\mathbf{x},t+1}^{\alpha,\eta}=\sum_{\alpha^{\prime}=1}^{N}\sum_{\eta^{%
\prime}=\pm}C_{\alpha^{\prime},\eta^{\prime}}^{\alpha,\eta}\psi_{\mathbf{x}%
-\eta\mathbf{u}_{\alpha},t}^{\alpha^{\prime},\eta^{\prime}}.  \label{map_x}
\end{equation}
Equation (\ref{map_x}) is the $N$-dimensional QW map in position representation. It shows that for any given time step
the wave-function at each point is the coherent linear superposition of the wave-functions at the neighboring points
calculated in the previous time step, the weights of the
superposition being given by the coin operator matrix elements $%
C_{\alpha^{\prime},\eta^{\prime}}^{\alpha,\eta}$.

Given the linearity of the map and the fact that it is space-invariant, i.e. the matrix elements
$C_{\alpha^{\prime},\eta^{\prime}}^{\alpha,\eta}$ do not depend on the space coordinates, the spatial Discrete Fourier
Transform (DFT), which has been used many times in QW studies \cite{Grimmett,nayak}, is a very useful technique.

The DFT is defined as
\begin{equation}
\left\vert \tilde{\psi}_{\mathbf{k,}t}\right\rangle \equiv\sum_{\mathbf{x}%
}e^{-i\mathbf{k}\cdot\mathbf{x}}\left\vert \psi_{\mathbf{x},t}\right\rangle ,
\label{DFT_k} \\
\end{equation}
where $\mathbf{k}=\left(k_{1},\ldots,k_{N}\right)$; $k_{\alpha}\in\left[%
-\pi,\pi\right]$, is the quasi-momentum vector. The DFT satisfies
\begin{equation}
\left\vert \psi_{\mathbf{x},t}\right\rangle \equiv\int\frac{\mathrm{d}^{N}%
\mathbf{k}}{\left(2\pi\right)^{N}}e^{i\mathbf{k}\cdot\mathbf{x}}\left\vert
\tilde{\psi}_{\mathbf{k},t}\right\rangle .  \label{DFT_x}
\end{equation}
Following Eq.(\ref{psi_x}) we define the components of the wavefunction in
momentum space as
\begin{align}
\left\vert \tilde{\psi}_{\mathbf{k},t}\right\rangle &
=\sum_{\alpha=1}^{N}\sum_{\eta=\pm}\tilde{\psi}_{\mathbf{k}%
,t}^{\alpha,\eta}\left\vert \alpha_{\eta}\right\rangle , \\
\tilde{\psi}_{\mathbf{k},t}^{\alpha,\eta} & =\sum_{\mathbf{x}}e^{-i\mathbf{k}%
\cdot\mathbf{x}}\psi_{\mathbf{x},t}^{\alpha,\eta}.
\end{align}
Applying the previous definitions to the map (\ref{map_x}), and using
\begin{equation}
\sum_{\mathbf{x}}e^{-i\mathbf{k}\cdot\mathbf{x}}\left\vert \psi_{\mathbf{x}%
-\eta\mathbf{u}_{\alpha},t}\right\rangle =\exp\left({-i\eta\mathbf{k}\cdot%
\mathbf{u}_{\alpha}}\right)\left\vert \tilde{\psi}_{\mathbf{k}%
,t}\right\rangle ,  \label{trasla}
\end{equation}
we obtain
\begin{equation}
\left\vert \tilde{\psi}_{\mathbf{k},t+1}\right\rangle =\hat{C}_{\mathbf{k}%
}\left\vert \tilde{\psi}_{\mathbf{k},t}\right\rangle ,  \label{mapket_k}
\end{equation}
where we have defined a coin operator in momentum space
\begin{equation}
\hat{C}_{\mathbf{k}}\equiv\sum_{\alpha=1}^{N}\sum_{\eta=\pm}\left\vert
\alpha_{\eta}\right\rangle \left\langle \alpha_{\eta}\right\vert \hat{C}%
\exp\left({-i\eta k_{\alpha}}\right).  \label{Ck}
\end{equation}
Above, $k_{\alpha}=\mathbf{k}\cdot\mathbf{u}_{\alpha}$.

The matrix elements of the coin operator in this space are
\begin{equation}
\left\langle \alpha_{\eta}\right\vert \hat{C}_{\mathbf{k}}\left\vert
\alpha_{\eta^{\prime}}^{\prime}\right\rangle \equiv\left(C_{\mathbf{k}%
}\right)_{\alpha^{\prime},\eta^{\prime}}^{\alpha,\eta}=\exp\left({-i\eta
k_{\alpha}}\right)C_{\alpha^{\prime},\eta^{\prime}}^{\alpha,\eta}.
\label{Ck_elements}
\end{equation}
Projecting Eq.(\ref{mapket_k}) onto $\left\langle \alpha_{\eta}\right\vert $
and using (\ref{Ck},\ref{Ck_elements}) leads to
\begin{equation}
\tilde{\psi}_{\mathbf{k},t+1}^{\alpha,\eta}=\sum_{\alpha^{\prime}=1}^{N}%
\sum_{\eta^{\prime}=\pm} \exp\left({-i\eta k_{\alpha}}%
\right)C_{\alpha^{\prime},\eta^{\prime}}^{\alpha,\eta} \tilde{\psi}_{\mathbf{%
k},t}^{\alpha^{\prime},\eta^{\prime}}.  \label{map_k}
\end{equation}
As we see, the nonlocal maps (\ref{mapket_x},\ref{map_x}) become local in
the momentum representation given by Eqs.(\ref{mapket_k}),(\ref{map_k}).
This allows us to easily obtain a formal solution to the QW dynamics, since
map (\ref{mapket_k}) implies
\begin{equation}
\left\vert \tilde{\psi}_{\mathbf{k},t}\right\rangle =\left(\hat{C}_{\mathbf{k%
}}\right)^{t}\left\vert \tilde{\psi}_{\mathbf{k},0}\right\rangle .
\label{evol_k}
\end{equation}
Therefore the set of eigenvalues and eigenvectors of $\hat{C}_{\mathbf{k}}$
is most useful to solve the QW evolution dynamics.

Since, according to Eq.(\ref{evol_k}) the operator $\hat{C}_{\mathbf{k}}%
\mathcal{\ }$ must be unitary, all its eigenvalues $\left\{ \lambda_{\mathbf{%
k}}^{\left(s\right)}: s=1,2,3,..,2N\right\}$ can be written in the form $%
\lambda_{\mathbf{k}}^{\left(s\right)}=\exp\left(-i\omega_{\mathbf{k}%
}^{\left(s\right)}\right)$, with $\omega_{\mathbf{k}}^{\left(s\right)}$
real. In addition to these eigenvalues we also need to know the
corresponding eigenvectors $\left\{\left\vert \phi_{\mathbf{k}%
}^{\left(s\right)}\right\rangle \right\}$. These eigenvectors satisfy the
orthogonality condition
\begin{equation}
\left\langle \phi_{\mathbf{k}}^{\left(s\right)}\right.\left\vert\phi_{%
\mathbf{k}}^{\left(s^{\prime }\right)}\right\rangle=\delta_{ss^{\prime }},
\label{or}
\end{equation}
where $\delta_{ss^{\prime }}$ is the Kronecker delta. Once the eigenvalues
and eigenvectors of $\hat{C}_{\mathbf{k}}$ are known, implementing Eq.(\ref%
{evol_k}) is straightforward. Given the initial distribution of the walker
in position representation $\left\vert \psi_{\mathbf{x},0}\right\rangle $,
we compute its DFT $\left\vert \tilde{\psi}_{\mathbf{k},0}\right\rangle $
via Eq.(\ref{DFT_k}), as well as the projections
\begin{equation}
\tilde{f}_{\mathbf{k}}^{\left(s\right)}=\left\langle \phi_{\mathbf{k}%
}^{\left(s\right)}\right.\left\vert \tilde{\psi}_{\mathbf{k},0}\right\rangle
,  \label{fsk}
\end{equation}
so that $\left\vert \tilde{\psi}_{\mathbf{k},0}\right\rangle =$ $\sum_{s}%
\tilde{f}_{\mathbf{k}}^{\left(s\right)}\left\vert \phi_{\mathbf{k}%
}^{\left(s\right)}\right\rangle $. Using Eq.(\ref{evol_k}), we obtain
\begin{equation}
\left\vert \tilde{\psi}_{\mathbf{k},t}\right\rangle =\sum_{s=1}^{2N}\exp{%
\left(-i\omega_{\mathbf{k}}^{\left(s\right)}t\right)}\tilde{f}_{\mathbf{k}%
}^{\left(s\right)}\left\vert \phi_{\mathbf{k}}^{\left(s\right)}\right\rangle
.
\end{equation}
In position representation we get, using Eq.(\ref{DFT_x}),
\begin{align}
\left\vert \psi_{\mathbf{x},t}\right\rangle & = \sum_{s=1}^{2N}\left\vert
\psi_{\mathbf{x},t}^{\left(s\right)}\right\rangle ,  \label{evol_x} \\
\left\vert \psi_{\mathbf{x},t}^{\left(s\right)}\right\rangle & =\int\frac{%
\mathrm{d}^{N}\mathbf{k}}{\left(2\pi\right)^{N}}\exp\left[{i\left(\mathbf{k}%
\cdot\mathbf{x-} \omega_{\mathbf{k}}^{\left(s\right)}t\right)}\right]\tilde{f%
}_{\mathbf{k}}^{\left(s\right)}\left\vert \phi_{\mathbf{k}%
}^{\left(s\right)}\right\rangle .  \label{evol_x_s}
\end{align}
In this way the time evolution of the QW is formally solved: all we need is
to compute the set of eigenvalues and eigenstates of $\hat{C}_{\mathbf{k}}$
and the initial state in reciprocal space $\left\vert \tilde{\psi}_{\mathbf{k%
},0}\right\rangle $, which determines the weight functions $\tilde{f}_{%
\mathbf{k}}^{\left(s\right)}$ through Eq.(\ref{fsk}).

\section{Entanglement and thermodynamics.}\label{thermo}

Entanglement in quantum mechanics is associated with the non separability of
the degrees of freedom of two or more particles. The degrees of freedom
involved in entangled states are usually discrete, such as the spins of
electrons or nuclei. However, there is also interest in continuous degrees
freedom, such as the position or the moment of a particle, due to their
potential to increase storage capacity and information processing in quantum
computation \cite{Malena}. The unitary evolution of the QW generates
entanglement between the coin and position degrees of freedom. The
asymptotic coin-position entanglement and its dependence on the initial
conditions of the QW has been investigated by several authors \cite%
{Carneiro,abal,salimi,Annabestani,Omar,Pathak,Petulante,Venegas,Endrejat,Ellinas1,Ellinas2,Maloyer,alejo2010,alejo2012}%
. In particular in Ref.\cite{alejo2012} it has been shown that the coin-position entanglement can be seen as a
system-environment entanglement and it allows to define an entanglement temperature. In the present work we also study
this subject using the $N$-dimensional QW as a system.

Let us briefly review the usual definition of entropy with the aim to
clarify the emergence of the concept of entanglement entropy. The density
matrix of the quantum system is
\begin{equation}
\widehat{\rho}(t)=\left\vert \psi_{t}\right\rangle \left\langle \psi_{t}
\right\vert .  \label{rho}
\end{equation}
The quantum analog of the Gibbs entropy is the von Neumann entropy
\begin{equation}
S _{N}(t)=-\mathrm{tr}(\widehat{\rho}(t) \log{\widehat{\rho}(t)}).
\label{sn}
\end{equation}
Owing to the unitary dynamics of the QW, the system remains in a pure state,
and this entropy vanishes. However, for these pure states, the entanglement
between the chirality and the position can be quantified by the associated
von Neumann entropy for the reduced density operator, namely
\begin{equation}
S(t)=-\mathrm{tr}(\widehat{\rho} _{c}(t) \log{\widehat{\rho} _{c}(t)}),
\label{s2}
\end{equation}
where
\begin{equation}
\widehat{\rho}_c(t)=\mathrm{tr_p}(\widehat{\rho} )=\sum_{\mathbf{x}%
}\left\langle \mathbf{x}\left\vert \psi_{t}\right\rangle \left\langle
\psi_{t} \right\vert \mathbf{x}\right\rangle ,  \label{rhoc}
\end{equation}
is the reduced density operator for the chirality evolution and the partial
trace, $\mathrm{tr_p}$, is taken over the positions. Note that, in general $%
\mathrm{tr}(\widehat{\rho}_{c}^{2})<1$, i.e., the reduced operator $\widehat{%
\rho}_{c}(t)$ corresponds to a statistical mixture. The expression for the
entropy given by Eq.(\ref{s2}), will be used as a measure of entanglement
between the position and the chirality of the system. Using the properties
of the wave-function $\left\vert \psi_{\mathbf{x},t}\right\rangle
=\left\langle \mathbf{x}\right.\left\vert \psi_{t}\right\rangle$ and the
identity
\begin{equation}
\sum_{\mathbf{x}}e^{i(\mathbf{k}-\mathbf{k}_{0})\cdot\mathbf{x}%
}=(2\pi)^{N}\delta^{N}\left(\mathbf{k}-\mathbf{k}_{0}\right),
\end{equation}
for the $N$-dimensional delta, it is straightforward to obtain the following
expression for Eq.(\ref{rhoc}), the reduced density operator

\begin{align}
\widehat{\rho}_c(t)& = \sum_{s=1}\sum_{s^{,}=1}\int\exp\left[{i
\left(\omega_{\mathbf{k}}^{\left(s^{,}\right)}-\omega_{\mathbf{k}%
}^{\left(s\right)}\right)t}\right]  \notag \\
& \times \tilde{f}_{\mathbf{k}}^{\left(s\right)} (\tilde{f}_{\mathbf{k}%
}^{\left(s^{,}\right)})^{*} \left\vert \phi_{\mathbf{k}}^{\left(s\right)}%
\right\rangle \left\langle\phi_{\mathbf{k}}^{\left(s^{,}\right)}\right\vert
\frac{\mathrm{d}^{N}\mathbf{k}}{\left(2\pi\right)^{N}}.  \label{evolrho}
\end{align}
This expression can be evaluated in the asymptotic limit $\mathrm{t}%
\rightarrow\infty$ using the stationary phase theorem, see Ref. \cite{nayak}%
, where only terms with $\omega_{\mathbf{k}}^{\left(s^{,}\right)}=\omega_{%
\mathbf{k}}^{\left(s\right)}$ contribute in Eq.(\ref{evolrho}). Therefore,
in the asymptotic limit the reduced density operator is
\begin{equation}
\widehat{\varrho}\equiv\widehat{\rho}_c(t\rightarrow\infty)=\sum_{s=1}^{2N}%
\int\frac{\mathrm{d}^{N}\mathbf{k}}{\left(2\pi\right)^{N}} |\tilde{f}_{%
\mathbf{k}}^{\left(s\right)}|^{2} \left\vert \phi_{\mathbf{k}%
}^{\left(s\right)}\right\rangle \left\langle\phi_{\mathbf{k}%
}^{\left(s\right)}\right\vert.  \label{evolrhoinfo}
\end{equation}
As the density operator is positive definite, its associated matrix, Eq.(\ref%
{evolrhoinfo}), has real and positive eigenvalues. We let $%
\{\left\vert\Phi_{s}\right\rangle\}$ be the basis that makes diagonal this
matrix. Therefore, in this basis, the corresponding asymptotic density
matrix has the following simple shape.
\begin{equation}
{\varrho}_{ss^{^{\prime }}}=\Lambda_s~\delta_{ss^{^{\prime }}},  \label{dia}
\end{equation}
where $\Lambda_s\geq 0$ are the eigenvalues of the asymptotic density
matrix, that satisfy
\begin{equation}
\sum_{s=1}^{2N}\Lambda_s=1.  \label{evolrhoinf}
\end{equation}
In order to make a more complete description of this equilibrium in the
asymptotic limit, it is necessary to connect the eigenvalues of $\rho_c$
with an unknown associated Hamiltonian operator $H_c$. To obtain this
connection we shall use the quantum Brownian motion model of Ref.\cite{Kubo}%
. In this theory one considers that the entanglement between the system
associated with the chirality degrees of freedom, characterized by the
density matrix $\rho_c$, and those associated with the position degrees of
freedom, the lattice, is equivalent to the thermal contact between the
system and a thermal bath. In equilibrium
\begin{equation}
[H_c,\rho_{c}]=0,  \label{scho2}
\end{equation}
should be satisfied. As a consequence, in the asymptotic regime the density
operator $\rho_{c}$ is an explicit function of a time-independent
Hamiltonian operator. If we note by $\{\left\vert\Phi_{s}\right\rangle\}$
the set of eigenfunctions of the density matrix, the operators $H_{c}$ and $%
\rho_{c}$ are both diagonal in this basis. Therefore the eigenvalues $%
\Lambda _{s}$ depend on the corresponding eigenvalues of $H_c$. We denote this set of eigenvalues by $\{\epsilon_s\}$;
they can be interpreted as the possible values of the entanglement energy. This interpretation agrees with the fact that
$\Lambda_{s}$ is the probability that the system is in the eigenstate $\left\vert\Phi_{s}\right\rangle$.

To construct this connection, we note that Eq.(\ref{evolrhoinf}) together $%
0\le \Lambda_s$ imply that $0\le \Lambda_s\le 1$, therefore making possible
it to associate a Boltzmann-type probability to each $\Lambda_s$. In other
words, it is possible to associate, to each $\Lambda_s$, a virtual level of
energy $\epsilon_s$. The precise dependence between $\Lambda _{s}$ and $%
\epsilon_s$ is determined by the type of ensemble we construct. We propose
in the present work that this equilibrium can be made to correspond to a
quantum canonical ensemble. To do this, we define the following relation
\begin{equation}
\Lambda _{s}\equiv\frac{e^{-\beta\epsilon_s}}{\mathbb{Z}},  \label{lam20}
\end{equation}
where $\mathbb{Z}$ is the partition function of the system, that is
\begin{equation}
\mathbb{Z}\equiv\sum_{s=1}^{2N}e^{-\beta\epsilon_s},  \label{part}
\end{equation}
and the parameter $\beta$ can be put into correspondence with an
entanglement temperature
\begin{equation}
T\equiv\frac{1}{\kappa\beta},  \label{tem}
\end{equation}
where $\kappa$ is the Boltzmann constant. Since only the relative difference between energy eigenvalues has physical
significance, we consider the eigenvalues in decreasing order, and, without loss of generality, set
\begin{equation}
\epsilon_1=\epsilon ,  \label{eney0}
\end{equation}
\begin{equation}
\epsilon_{2N}=-\epsilon .  \label{eney}
\end{equation}
The value of $\epsilon$ can be determined from Eqs.(\ref{lam20},\ref{eney0},%
\ref{eney})
\begin{equation}
\epsilon=\frac{1}{2\beta}\log\frac{\Lambda _{2N}}{\Lambda _{1}} .
\label{eney00}
\end{equation}
The energy eigenvalues for the remaining values of s, $s=2,3...,2N-1$, are,
using again Eq.(\ref{lam20}),
\begin{equation}
\epsilon_{s}=\epsilon -\frac{1}{\beta}\log\frac{\Lambda _{s}}{\Lambda _{1}}. \label{eneres}
\end{equation}
Therefore the asymptotic density matrix of Eq.(\ref{dia}) can be thought as
the density matrix of the canonical ensemble
\begin{equation}
{\varrho}=\frac{1}{\mathbb{Z}}
\begin{pmatrix}
e^{-\beta \epsilon_{1}} & 0 & 0 & . & . & 0 & 0 \\
0 & e^{-\beta \epsilon_{2}} & 0 & . & . & 0 & 0 \\
0 & 0 & e^{-\beta \epsilon_{3}} & . & . & 0 & 0 \\
. & . & . & . & . & . & . \\
. & . & . & . & . & . & . \\
0 & 0 & 0 & . & . & e^{-\beta \epsilon_{2N-1}} & 0 \\
0 & 0 & 0 & . & . & 0 & e^{-\beta \epsilon_{2N}}%
\end{pmatrix}%
.  \label{dia2}
\end{equation}

Starting from the partition function of the system given by Eq.(\ref{part}),
it is possible to build the thermodynamics for the QW entanglement. In
particular, the Helmholtz free energy $A$ is given by
\begin{equation}
A\equiv-\frac{1}{\beta}\log\mathbb{Z}=-\frac{1}{\beta}\log%
\sum_{s=1}^{2N}e^{-\beta\epsilon_s}.  \label{free}
\end{equation}
and the internal energy $U$ is given by
\begin{equation}
U\equiv-\frac{1}{\mathbb{Z}}\frac{\partial\mathbb{Z}}{\partial\beta}=\frac{1%
}{\mathbb{Z}}\sum_{s=1}^{2N}\epsilon_s e^{-\beta\epsilon_s}.  \label{free2}
\end{equation}
Thus, the asymptotic entanglement entropy as a function of the eigenvalues $%
\Lambda _{s}$ is
\begin{equation}
S=-\sum_{s=1}^{2N}\Lambda _{s} \log{\Lambda _{s}}.  \label{s00}
\end{equation}
Substituting Eq.(\ref{lam20}) into Eq.(\ref{s00}), after straightforward
operations using Eqs.(\ref{free},\ref{free2}), we obtain the following
expression for the asymptotic entanglement entropy
\begin{equation}
{S}= {\beta}(U- A).  \label{termo}
\end{equation}
As it should be expected, this last equation agrees with the thermodynamic
definition of the entropy.

Of course, in Eq.(\ref{eney00}) only the ratio $\epsilon/T$ is well defined; however, we chose to introduce the
temperature as this concept strengthens the idea of asymptotic equilibrium between the position and chirality degrees of
freedom. Note that while temperature makes sense only in the mentioned equilibrium state, the entropy concept can be
introduced without such a restriction. For all practical purposes we shall take $\epsilon=\kappa$, then the entanglement
temperature will be determined by
\begin{equation}
T=\frac{2}{\log\left({\Lambda _{2N}}/{\Lambda _{1}}\right)},  \label{tem00}
\end{equation}
and the energy eigenvalues by
\begin{equation}
\epsilon_{s}=1-2\frac{\log\left({\Lambda _{s}}/{\Lambda _{1}}\right)}{%
\log\left({\Lambda _{2N}}/{\Lambda _{1}}\right)}.  \label{enefin}
\end{equation}

\section{Initial conditions.}\label{initial}

We now discuss the consequences of choosing different initial conditions on
the thermal evolution of the system. We are interested in characterizing the
long-time coin-position entanglement generated by the evolution of the $N$%
-dimensional QW. First we consider the case of a separable coin-position
initial state. More specifically, we take initial chirality conditions of
the form
\begin{equation}
\left\vert {\psi _{\mathbf{x},0}}\right\rangle ={\xi _{\mathbf{x},0}}%
\left\vert {\chi }\right\rangle ,  \label{inipsi3}
\end{equation}%
where ${\xi _{\mathbf{x},0}}$  is a generic position wave function and
\begin{equation}
\left\vert {\chi }\right\rangle =\cos {(\gamma /2)}\left\vert {Z_+ }%
\right\rangle +e^{i\varphi }\sin {(\gamma /2)}\left\vert {Z_- }\right\rangle,
\label{inipsi4}
\end{equation}
with
\begin{equation}
\left\vert {Z_\pm }\right\rangle \equiv\frac{1}{\sqrt{N}}\sum_{\alpha =1}^{N}\left\vert \ \alpha _{\pm}\right\rangle .
\label{z1}
\end{equation}
The two parameters $\gamma \in \left[ 0,\pi \right] $ and $\varphi \in\left[
0,2\pi \right] $ define the initial point on the generalized Bloch's sphere. %
The DFT of Eq.(\ref{inipsi3}) is
\begin{equation}
\left\vert \tilde{\psi}_{\mathbf{k,}0}\right\rangle =\sum_{\mathbf{x}}e^{-i%
\mathbf{k}\cdot \mathbf{x}}\left\vert \psi _{\mathbf{x},0}\right\rangle
=\sum_{\mathbf{x}}e^{-i\mathbf{k}\cdot \mathbf{x}}\xi _{\mathbf{x}%
,0}\left\vert {\chi }\right\rangle .  \label{DFT_k0}
\end{equation}%
In order to obtain a closed equation for $\Lambda _{s}$ we consider in
detail the simple case where the amplitudes ${\xi _{\mathbf{x},0}}$ have an
isotropic Gaussian position distribution multiplied by the plane waves $e^{i%
\mathbf{k_{0}}\cdot \mathbf{x}}$ , that is
\begin{equation}
{\xi _{\mathbf{x},0}}\propto e^{i\mathbf{k_{0}}\cdot \mathbf{x}}\frac{1}{%
\sigma ^{N/2}}\exp {\left( -\frac{\mathbf{x}\cdot \mathbf{x}}{\sigma ^{2}}%
\right) }.  \label{gauss1}
\end{equation}%
where $\sigma >0$ is a characteristic width and $\mathbf{k_{0}}$ is a
particular initial momentum that characterized the initial condition. We
will deal with sufficiently large value of $\sigma $ for the Gaussian, so as
to make possible the connection of the DFT with the continuous limit. Then,
for these values of $\sigma $, Eq.(\ref{DFT_k0}) can be put as
\begin{equation}
\left\vert \tilde{\psi}_{\mathbf{k,}0}\right\rangle \propto \sigma
^{N/2}\sum_{\mathbf{x}}e^{-\frac{\sigma ^{2}}{2}{\left( \mathbf{k-k_{0}+2\pi
\mathbf{x}}\right) ^{2}}}\left\vert {\chi }\right\rangle ,  \label{gauss2}
\end{equation}%
see Appendix A. If we want to simulate an uniform initial distribution for the $N$-dimensional QW we can take $\sigma
\mapsto \infty $ in Eq.(\ref{gauss2}). In this case we can use the following mathematical property for the Dirac delta,
\begin{equation}
\lim_{\sigma \mapsto \infty }\left( \frac{\sigma }{\sqrt{\pi }}\right)
^{N}e^{-\sigma ^{2}{\left( \mathbf{k-k_{0}+2\pi \mathbf{x}}\right) ^{2}}%
}\equiv \delta ^{N}\left( \mathbf{k-k_{0}+2\pi \mathbf{x}}\right) .
\label{fsk3}
\end{equation}%
Eq.(\ref{gauss2}) can then be expressed as
\begin{equation}
\left\vert \tilde{\psi}_{\mathbf{k,}0}\right\rangle \propto \left[ \sum_{%
\mathbf{x}}\delta ^{N/2}\left( \mathbf{k-k_{0}+2\pi \mathbf{x}}\right) %
\right] \left\vert {\chi }\right\rangle .  \label{gauss3}
\end{equation}%
We shall now assume that the components of $\mathbf{k_{0}}$ belong to the
interval $\left( -\pi ,\pi \right) $, then in the sum of Eq.(\ref{gauss3})
the only term that survives is the one for $\mathbf{x}=\mathbf{0}$. This is
due to the fact that all components of $\mathbf{k}$ lie within the interval $%
\left[ -\pi ,\pi \right] $, and that the vector $\mathbf{x}$ has only
discrete components. Then using Eq.(\ref{fsk}), Eq.(\ref{gauss3}) and the
normalization condition, we have
\begin{equation}
\left\vert \tilde{f}_{\mathbf{k}}^{\left( s\right) }\right\vert ^{2}={\left(
2\pi \right) ^{N}}\delta ^{N}\left( \mathbf{k-k_{0}}\right) \left\vert
\left\langle \phi _{\mathbf{k}}^{\left( s\right) }\right. \left\vert \chi
\right\rangle \right\vert ^{2}.  \label{fsk4}
\end{equation}%
Therefore in this case, from Eq.(\ref{evolrhoinfo}), it is straightforward to
obtain the eigenvalues for the asymptotic density matrix,
\begin{equation}
\Lambda _{s}={\left\vert \left\langle \phi _{\mathbf{k_{0}}}^{\left(
s\right) }\right. \left\vert \chi \right\rangle \right\vert ^{2}},
\label{lam2}
\end{equation}%
and their respective eigenfunctions,
\begin{equation}
\left\vert \Phi _{s}\right\rangle =\left\vert \phi _{\mathbf{k_{0}}}^{\left(
s\right) }\right\rangle .  \label{eigen}
\end{equation}

As a second example we consider the case of a non separable coin-position
initial state. In particular we take
\begin{equation}
\left\vert \psi _{\mathbf{x},0}\right\rangle =\int \frac{\mathrm{d}^{N}%
\mathbf{k}}{\left( 2\pi \right) ^{N}}\exp \left[ {i\left( \mathbf{k}\cdot
\mathbf{x}\right) }\right] \left\vert \tilde{\psi}_{\mathbf{k,}%
0}\right\rangle ,  \label{inidos}
\end{equation}%
with
\begin{align}
\left\vert \tilde{\psi}_{\mathbf{k,}0}\right\rangle & =\cos {(\gamma /2)}%
\frac{1}{\sqrt{N}}\sum_{s=1}^{N}\left\vert {\phi _{\mathbf{k}}^{\left(
s\right) }}\right\rangle  \notag \\
& +e^{i\varphi }\sin {(\gamma /2)}\frac{1}{\sqrt{N}}\sum_{s=N+1}^{2N}\left%
\vert {\phi _{\mathbf{k}}^{\left( s\right) }}\right\rangle ,  \label{renato}
\end{align}%
and then
\begin{eqnarray}  \label{renato2}
\left\vert \tilde{f}_{\mathbf{k}}^{\left( s\right) }\right\vert ^{2}
&=&\left\vert \left\langle \phi _{\mathbf{k}}^{\left( s\right) }\right.
\left\vert \tilde{\psi}_{\mathbf{k},0}\right\rangle \right\vert ^{2}  \notag
\\
&=&\frac{1}{N}\left\{
\begin{array}{c}
\cos ^{2}{(\gamma /2)}, \\
\sin ^{2}{(\gamma /2),}%
\end{array}%
\begin{array}{c}
\text{for }s=1,2,...,N \\
\notag \text{for }s=N+1,N+2,...,2N%
\end{array}%
\right. \\
\end{eqnarray}%
Therefore the eigenvalues $\Lambda _{s}$ are the eigenvalues of the matrix
associated to the following operator, see Eq.(\ref{evolrhoinfo})
\begin{eqnarray}
&&\frac{1}{N}\int \frac{\mathrm{d}^{N}\mathbf{k}}{\left( 2\pi \right) ^{N}}%
\left\{ \cos ^{2}{(\gamma /2)}\sum_{s=1}^{N}\left\vert \phi _{\mathbf{k}%
}^{\left( s\right) }\right\rangle \left\langle \phi _{\mathbf{k}}^{\left(
s\right) }\right\vert \right.  \notag \\
&&+\sin ^{2}{(\gamma /2)}\left. \sum_{s=N+1}^{2N}\left\vert \phi _{\mathbf{k}%
}^{\left( s\right) }\right\rangle \left\langle \phi _{\mathbf{k}}^{\left(
s\right) }\right\vert \right\} .  \label{renato3}
\end{eqnarray}%

As a third example, we take 
\begin{align}
\left\vert \tilde{\psi}_{\mathbf{k,}0}\right\rangle & =\cos {(\gamma /2)}%
\frac{1}{\sqrt{N}}\sum_{s=1}^{N}\left\vert {\phi _{\mathbf{k}}^{\left(
2s\right) }}\right\rangle  \notag \\
& +e^{i\varphi }\sin {(\gamma /2)}\frac{1}{\sqrt{N}}\sum_{s=1}^{N}\left\vert
{\phi _{\mathbf{k}}^{\left( 2s-1\right) }}\right\rangle ,  \label{iniuno}
\end{align}%
and then
\begin{equation}
\left\vert \tilde{f}_{\mathbf{k}}^{\left( s\right) }\right\vert
^{2}=\left\vert \left\langle \phi _{\mathbf{k}}^{\left( s\right) }\right.
\left\vert \tilde{\psi}_{\mathbf{k},0}\right\rangle \right\vert ^{2}=\frac{1%
}{N}\left\{
\begin{array}{c}
\cos ^{2}{(\gamma /2)}, \\
\sin ^{2}{(\gamma /2),}%
\end{array}%
\begin{array}{c}
\text{for }s\text{ even} \\
\text{for }s\text{ odd}%
\end{array}%
\right. .  \label{fsk2}
\end{equation}%
Finally, using Eq.(\ref{evolrhoinfo}), the eigenvalues $\Lambda _{s}$ are
the eigenvalues of the matrix associated to the operator
\begin{eqnarray}
&&\frac{1}{N}\int \frac{\mathrm{d}^{N}\mathbf{k}}{\left( 2\pi \right) ^{N}}%
\left\{ \cos ^{2}{(\gamma /2)}\sum_{s=1}^{N}\left\vert \phi _{\mathbf{k}%
}^{\left( 2s\right) }\right\rangle \left\langle \phi _{\mathbf{k}}^{\left(
2s\right) }\right\vert \right.  \notag \\
&&+\sin ^{2}{(\gamma /2)}\left. \sum_{s=1}^{N}\left\vert \phi _{\mathbf{k}%
}^{\left( 2s-1\right) }\right\rangle \left\langle \phi _{\mathbf{k}}^{\left(
2s-1\right) }\right\vert \right\} .  \label{lam0}
\end{eqnarray}

\section{Application to the 2D quantum walk}\label{examples2D}

In this Section we illustrate the general treatment introduced above in the
special case of the $2D$ quantum walk. References \cite{Inui,Watabe08}
introduced a one-parameter family of quantum-walk models on $2D$ as a
generalization of Grover's model by specifying the corresponding matrix $C_{%
\mathbf{k}}$, see Eq.(\ref{Ck_elements}), as
\begin{equation}
C_{\mathbf{k}}=%
\begin{pmatrix}
-pe^{ik_{1}} & qe^{ik_{1}} & \sqrt{pq}e^{ik_{1}} & \sqrt{pq}e^{ik_{1}} \\
qe^{-ik_{1}} & -pe^{-ik_{1}} & \sqrt{pq}e^{-ik_{1}} & \sqrt{pq}e^{-ik_{1}}
\\
\sqrt{pq}e^{ik_{2}} & \sqrt{pq}e^{ik_{2}} & -qe^{ik_{2}} & pe^{ik_{2}} \\
\sqrt{pq}e^{-ik_{2}} & \sqrt{pq}e^{-ik_{2}} & pe^{-ik_{2}} & -qe^{-ik_{2}}%
\end{pmatrix}%
,  \label{G_k}
\end{equation}
where the parameter $p\in[0,1]$, $q=1-p$ and $\mathbf{k}=\left(k_{1},k_{2}%
\right)$ is the quasi-momentum vector. If $p=q=1/2$ we have the Grover coin.
From now on we take this to be the case.

Eq.(\ref{G_k}) has four eigenvalues $\lambda_{s},\;s=1,2,3,4$,
\begin{align}
\lambda_{1}=1,\lambda_{2}=-1,\lambda_{3}=e^{i\omega\left(k_{1},k_{2}%
\right)},\lambda_{4}=e^{-i\omega\left(k_{1},k_{2}\right)},  \label{auto}
\end{align}
where
\begin{equation}
\cos\omega\left(k_{1},k_{2}\right)=-\frac{1}{2}\left(\cos{k_{1}}+\cos{k_{2}}%
\right).  \label{dis}
\end{equation}
The eigenvectors corresponding to the eigenvalues are given by the following
column vectors
\begin{equation}
\left\vert \phi_{\mathbf{k}}^{\left(s\right)} \right\rangle=\frac{1}{%
\mathcal{N}_{\mathbf{k}}^{(s)}}
\begin{pmatrix}
\left(1+e^{-\mathrm{i}k_{1}}\lambda_{\mathbf{k}}^{\left(s\right)}\right)^{-1}
\\
\left(1+e^{+\mathrm{i}k_{1}}\lambda_{\mathbf{k}}^{\left(s\right)}\right)^{-1}
\\
\left(1+e^{-\mathrm{i}k_{2}}\lambda_{\mathbf{k}}^{\left(s\right)}\right)^{-1}
\\
\left(1+e^{+\mathrm{i}k_{2}}\lambda_{\mathbf{k}}^{\left(s\right)}\right)^{-1}%
\end{pmatrix}%
,  \label{eig_gen}
\end{equation}
where the normalization factors $\mathcal{N}_{\mathbf{k}}^{(s)}$ are given
by
\begin{align}
\mathcal{N}_{\mathbf{k}}^{(1)}=\sqrt{\frac{1}{1+\cos{k_1}}+\frac{1}{1+\cos{%
k_2}}}  \notag \\
\mathcal{N}_{\mathbf{k}}^{(2)}=\sqrt{\frac{1}{1-\cos{k_1}}+\frac{1}{1-\cos{%
k_2}}}  \notag \\
\mathcal{N}_{\mathbf{k}}^{(3)}=\mathcal{N}_{\mathbf{k}}^{(4)}= \sqrt{2\,%
\frac{4-\left(\cos{k_1}+\cos{k_2}\right)^2}{\left(\cos{k_1}-\cos{k_2}%
\right)^2}}.  \label{norm}
\end{align}

From Eq.(\ref{auto}), we see that the first two eigenvalues $\lambda_{1}=1$
and $\lambda_{2}=-1$ do not depend on $k$, and the last two eigenvalues are
complex conjugates of each other. Equation (\ref{dis}) is a dispersion
relation of the system. The frequency $\omega\left(k_{1},k_{2}\right)\in%
\left[0,2\pi\right]$ and when $k_{1}=0$ and $k_{2}=0$ the system has a
degeneracy because the three eigenvalues $\lambda_{2}=\lambda_{3}=%
\lambda_{4}=-1$, see Eqs.(\ref{auto}, \ref{dis}). Then, due to this
degeneracy the frequencies $\pm\omega\left(k_{1},k_{2}\right)$, as a function
of $k_{1}$ and $k_{2}$, has a diabolo shape. These degenerate points are
called ``diabolical points" \cite{german2013}.
\begin{figure}[!ht]
\centering
\includegraphics[width=0.7\columnwidth]{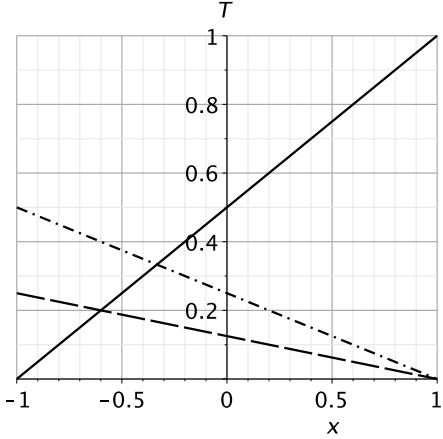}
\caption{The eigenvalues of the reduced density matrix, Eqs.(\protect\ref%
{lam1},\protect\ref{lam2},\protect\ref{lam34}), as a function of the
parameter $x=\sin{\protect\gamma}\cos{\protect\varphi}$, with $\protect\theta%
=\protect\pi$. $\Lambda_1$ in full line, $\Lambda_2$ in dashed line and $%
\Lambda_3$ in dot-dashed line.} \label{fig:a}
\end{figure}
\begin{figure}[!ht]
\centering
\includegraphics[width=0.7\columnwidth]{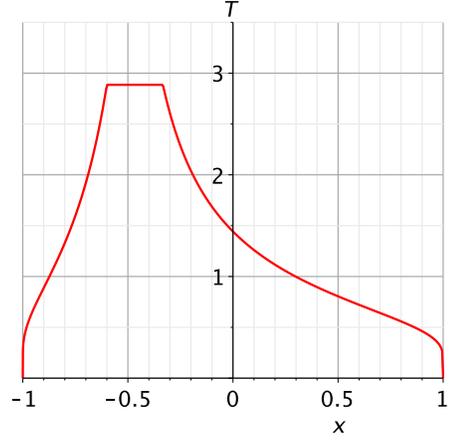}
\caption{Entanglement temperature, see Eq.(\protect\ref{temk0}), as function
of the dimensionless parameter $x=\sin{\protect\gamma}\cos{\protect\varphi}$%
, with $\protect\theta=\protect\pi$.} \label{fig:b}
\end{figure}
\subsection{QW's temperature for a separable coin-position initial state}\label{exampleA}

In order to calculate $\Lambda_s$, Eq.(\ref{lam2}), we select the diabolical
point $\mathbf{k_0}=\mathbf{0}$ and we must be very careful, because the
calculation of the eigenvectors, Eq.(\ref{eig_gen}), has indeterminacies.
The eigenvectors of the $2D$ Grover walk matrix are given by Eq.(\ref%
{eig_gen}). Whenever $\mathbf{k}$ is not close to a diabolical point these
eigenvectors vary smoothly around $\mathbf{k}$. However, we want to study
the behavior of the eigenvectors close to the diabolical point at $\mathbf{k}%
=\mathbf{k}_{\mathrm{0}}\equiv\left(0,0\right)$. We find it convenient to
use polar coordinates $\left(k_{1},k_{2}\right)=\left(k\cos\theta,k\sin%
\theta\right)$. Performing the limit of (\ref{eig_gen}) for $k\rightarrow0$
we find
\begin{align}
\left\vert\phi_{\mathbf{k_0}}^{\left(1\right)}\right\rangle & =\frac{1}{2}%
\left(%
\begin{array}{c}
1 \\
1 \\
1 \\
1%
\end{array}%
\right),  \label{eigen_close1} \\
\left\vert\phi_{\mathbf{k_0}}^{\left(2\right)}\right\rangle & =\frac{i}{%
\sqrt{2}}\left(%
\begin{array}{c}
-\sin\theta \\
+\sin\theta \\
-\cos\theta \\
+\cos\theta%
\end{array}%
\right),  \label{eigen_close2} \\
\left\vert\phi_{\mathbf{k_0}}^{\left(3\right)}\right\rangle & =\frac{i}{2%
\sqrt{2}}\left(%
\begin{array}{c}
1-\sqrt{2}\cos\theta \\
1+\sqrt{2}\cos\theta \\
-1+\sqrt{2}\sin\theta \\
-1-\sqrt{2}\sin\theta%
\end{array}%
\right),  \label{eigen_close3} \\
\left\vert\phi_{\mathbf{k_0}}^{\left(4\right)}\right\rangle & =\frac{i}{2%
\sqrt{2}}\left(%
\begin{array}{c}
-1-\sqrt{2}\cos\theta \\
-1+\sqrt{2}\cos\theta \\
1+\sqrt{2}\sin\theta \\
1-\sqrt{2}\sin\theta%
\end{array}%
\right).  \label{eigen_close4}
\end{align}
Taking the two-dimensional expression of $\left\vert{\chi}\right\rangle$,
see Eq.(\ref{inipsi4}), in its matrix shape
\begin{equation}
\left\vert{\chi}\right\rangle=\frac{1}{\sqrt{2}}
\begin{pmatrix}
\cos{(\gamma/2)} \\
e^{i\varphi}\sin{(\gamma/2)} \\
\cos{(\gamma/2)} \\
e^{i\varphi}\sin{(\gamma/2)}%
\end{pmatrix}%
,  \label{matini}
\end{equation}
we can evaluate $\Lambda_s$, see Eq.(\ref{lam2}), that is
\begin{equation}
\Lambda_1=\frac{1}{2}\left(1+\sin{\gamma}\cos{\varphi}\right),  \label{lam1}
\end{equation}
\begin{equation}
\Lambda_2=\frac{1}{4}\left(1+\sin{2\theta}\right)\left(1-\sin{\gamma}\cos{%
\varphi}\right),  \label{lam21}
\end{equation}
\begin{equation}
\Lambda_3=\Lambda_4=\frac{1}{8}\left(1-\sin{2\theta}\right)\left(1-\sin{%
\gamma}\cos{\varphi}\right).  \label{lam34}
\end{equation}
Figure~\ref{fig:a} shows the dependence of $\Lambda_s, s=1,2,3,4$ with the
initial conditions given through the parameter
\begin{equation}
x\equiv\sin{\gamma}\cos{\varphi}.  \label{xx}
\end{equation}

From Eq.(\ref{tem00}), the entanglement temperature in the diabolical point is
\begin{equation}
T={2/\log\left(\frac{\Lambda_{max}}{\Lambda_{min}} \right)},  \label{temk0}
\end{equation}
where $\Lambda_{max}$ and $\Lambda_{min}$ are respectively the maximum and minimum value of $\Lambda$ given by
Eqs.(\ref{lam1},\ref{lam21},\ref{lam34}).

Equation (\ref{temk0}) shows that the QW initial conditions $\gamma,\varphi$ and $\theta$ ($\mathbf{k_0}$) determine the
entanglement temperature and for a fixed $\theta$ the isothermal lines as a function of the initial conditions are
determined by the following equation
\begin{equation}
x=\sin{\gamma}\cos{\varphi}=\mathcal{C},  \label{iso1}
\end{equation}
where $\mathcal{C}$ is a constant.

In Fig.~\ref{fig:b} we see that the temperature as a function of $x$ increases from $T=0$ for $x=-1$ to the constant
value  $T_0=2/\log2$ in the $x$ interval $[-3/5,-1/3]$, and then decreases gradually, reaching $T=0$ at $x=1$. The
isotherms are the intersections of the Bloch sphere with the planes $x=constant$ .
\begin{figure}[!ht]
\centering
\includegraphics[width=0.7\columnwidth]{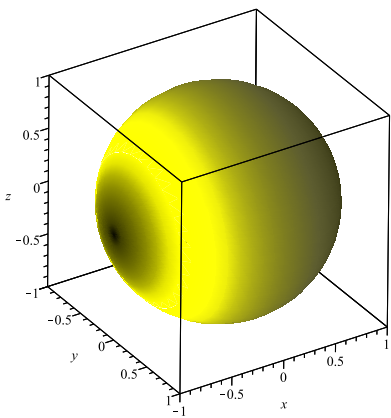}
\caption{{(Color online) Isotherms on the Bloch sphere. $\left\vert {Z_+ }\right\rangle$ and $\left\vert
{Z_-}\right\rangle$ are the North and South Pole, respectively. The two black points (``cold points", corresponding to
$T=0$) on the sphere are the points $\frac{1}{\sqrt{2}}(\left\vert {Z_+}\right\rangle+\left\vert {Z_- }\right\rangle)$
and $\frac{1}{\sqrt{2}}(\left\vert {Z_+ }\right\rangle-\left\vert {Z_-}\right\rangle)$. The light (yellow) zone is the
``hot zone" $T=T_0$.}} \label{fig:bloch}
\end{figure}
Figure~\ref{fig:bloch} shows the isotherms for the entanglement temperature  as a function of the QW initial position,
defined on the Bloch sphere. The figure shows three regions, two dark  zones left and right, corresponding to
temperatures $0<T<T_0$, and the a light one corresponding to the constant temperature $T=T_0$.

\subsection{QW's temperature for a non separable coin-position initial state I}\label{exampleB}
\begin{figure}[!ht]
\centering
\includegraphics[width=0.7\columnwidth]{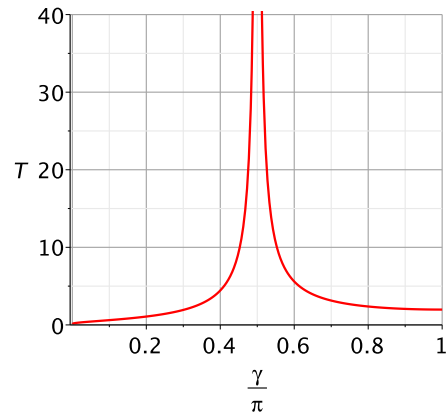}
\caption{Entanglement temperature, see Eq.~(\protect\ref{rena}), as a
function of the dimensionless parameter $\protect\gamma$.}
\label{fig:1}
\end{figure}

Taking the initial state given by Eqs.(\ref{inidos},\ref{renato}) and adding Eq.(\ref{renato3}), it is easy to show that
$\widehat{\varrho }$ reduces to
\begin{equation}
\widehat{\varrho }=\frac{1}{4}\left(
\begin{array}{cccc}
1 & a & b & b \\
\noalign{\medskip}a & 1 & b & b \\
\noalign{\medskip}b & b & 1 & a \\
\noalign{\medskip}b & b & a & 1%
\end{array}%
\right) ,  \label{renato11}
\end{equation}%
where%
\begin{eqnarray}
a &=&\left( 1-4/{\pi }\right) \cos \left( \gamma \right) ,  \label{re1} \\
b &=&\left( 1-2/{\pi }\right) \cos \left( \gamma \right) .  \label{re2}
\end{eqnarray}%
The eigenvalues of Eq.(\ref{renato11}) are
\begin{eqnarray}
\Lambda _{1} &=&[1-\cos (\gamma )]/4,  \label{r1} \\
\Lambda _{2} &=&[1-(3-8/\pi)\cos (\gamma )]/4,  \label{rr2} \\
\Lambda _{3} &=&[1-(1-4/\pi)\cos (\gamma )]/4\\
\Lambda _{4} &=&\Lambda _{3}.
\end{eqnarray}%
The entanglement temperature Eq.(\ref{tem00}) is thus given by
\begin{equation}
T=\frac{2}{\left\vert \ln  \frac{1+\left( \frac{4}{\pi }-1\right) \cos \gamma}{1-\cos \gamma}\right\vert}.  \label{rena}
\end{equation}%
Figure~\ref{fig:1} shows that the temperature as a function of $\gamma$ increases from $T=0$ for $\gamma=0$, to infinity
for $\gamma=\pi/2$, and then decreases gradually to $T={2}/{\left\vert \ln \left( 1-2/\pi \right) \right\vert }$ at
$\gamma=\pi$.
In order to take the initial condition on the generalized Bloch sphere, we redefine
\begin{equation}
\left\vert {Z_+ }\right\rangle \equiv \frac{1}{\sqrt{N}}\sum_{s=1}^{N}\left\vert {\phi _{\mathbf{k}}^{\left( s\right)
}}\right\rangle , \label{z1p}
\end{equation}
\begin{equation}
\left\vert {Z_- }\right\rangle \equiv \frac{1}{\sqrt{N}}\sum_{s=N+1}^{2N}\left%
\vert {\phi _{\mathbf{k}}^{\left( s\right) }}\right\rangle . \label{z2p}
\end{equation}
Then the initial state Eq.(\ref{renato}) takes the following form
\begin{align}
\left\vert \tilde{\psi}_{\mathbf{k,}0}\right\rangle  =\cos {(\gamma /2)}\left\vert {Z_+ }\right\rangle 
 +e^{i\varphi }\sin {(\gamma /2)} \left\vert {Z_- }\right\rangle,  \label{renato000000}
\end{align}
where $\gamma$ and $\varphi$ define a point on the unit Bloch sphere. In this case the isotherms have a rotation symmetry
around the axis defined by the points $\left\vert {Z_+}\right\rangle$ and $\left\vert {Z_-}\right\rangle$, North and
South poles respectively. Therefore the isotherms are the parallels $z = constant$ on the Bloch sphere. In the northern
hemisphere the temperatures of the isotherms increases from $T=0$, in the North pole, to infinity at the Equator, and on
the southern one the temperature of the isotherms decreases from infinity at the Equator, to the finite value
$T={2}/{\left\vert \ln \left( 1-2/\pi \right) \right\vert }$ in the South pole.

\subsection{QW's temperature for non separable coin-position initial state II}\label{exampleC}
For the 2D case, taking the initial state given by Eq.(\ref{iniuno}), after some heavy but straightforward operations, we
can evaluate $\Lambda _{s}$ and they satisfy
\begin{equation}
\Lambda _{s}=\frac{1}{4}, \mathrm{for}~~ s=1,2,3,4,   \label{cuatro1}
\end{equation}
which, according to Eq.(\ref{tem00}), indicates that the temperature is infinite all over the Bloch sphere, representing
a degenerate case. The symmetries of the Grover coin seem to point out that $\widehat{\varrho }=\frac{\hat{I}}{2N}$ for
$N>2$ when we use the initial condition Eq.(\ref{iniuno}).

\section{Conclusion}\label{conclusion}
During the last thirty years, several technological advances have made possible to construct and preserve quantum states.
They also have increased the possibility of building quantum computing devices. Therefore, the study of the dynamics of
open quantum systems becomes relevant both for development of these technologies as well as for the algorithms that will
run on those future quantum computers. The quantum walk has emerged as a useful theoretical tool to study many
fundamental aspects of quantum dynamics. It provides a frame to study, among other effects, the entanglement between its
degrees of freedom, in a simple setting that often allows for a full analytical treatment of the problem. The study of
this kind of entanglement is important in order to understand the asymptotic equilibrium between its internal degrees of
freedom.

In this paper we have studied the asymptotic regime of the $N$-dimensional quantum walk. We have focused into the
asymptotic entanglement between chirality and position degrees of freedom, and have shown that the system establishes a
stationary entanglement between the coin and the position that allows to develop a thermodynamic theory. Then we were
able to generalize previous results, obtained in references \cite{alejo2012,gustavo}. The asymptotic reduced density
operator was used to introduce the entanglement thermodynamic functions in the canonical equilibrium. These thermodynamic
functions characterize the asymptotic entanglement and the system can be seen as a particle coupled to an infinite bath,
the $|x\rangle$ position states. It was shown that the QW initial condition determines the system's temperature, as well
as other thermodynamic functions. A map for the isotherms was analytically built for arbitrary localized initial
conditions. The behavior of the reduced density operator looks diffusive but it has a dependence on the initial
conditions, the global evolution of the system being unitary. Then, if an observer only had information related with the
chirality degrees of freedom, it would be very difficult for it to recognize the unitary character of the quantum
evolution. In general, from this simple model we can conclude that if the quantum system dynamics occurs in a composite
Hilbert space, then the behavior of the operators that acts on only one sub-space could camouflage the unitary character
of the global evolution.

The development of experimental techniques has made possible the trapping of samples of atoms using resonant
exchanges in momentum and energy between atoms and laser light. However, it is not yet possible to prepare a
system with a particular initial chirality. Therefore, the average thermodynamical functions could have more
meaning when considered from an experimental point of view. It is interesting to point out that for a given
family of initial conditions, such as that given by Eq.(\ref{inipsi4}), the explicit dependence of thermodynamic
functions with the initial position on the Bloch's sphere, $\gamma$ and $\varphi$, can be eliminated if we take
the average of $\Lambda_s$ over all initial conditions. Then each family could be characterized by a single
asymptotic average temperature.

\

We acknowledge the support from PEDECIBA and ANII (FCE-2-211-1-6281, Uruguay), CNPq and LNCC (Brazil), and the
CAPES-UdelaR collaboration program. FLM acknowledges financial support from FAPERJ/APQ1, CNPq/Universal and CAPES/AUXPE
grants.

\appendix
\section{}
Here we derive Eq.(\ref{gauss2}). We employ the well known Poisson summation
formula
\begin{equation}
\sum_{n=-\infty}^{n=\infty}g(n)=
\sum_{n=-\infty}^{n=\infty}\int_{-\infty}^{\infty}g(x)e^{-i2\pi n x} dx,
\label{Poisson}
\end{equation}
which, together with Eqs.(\ref{DFT_k0},\ref{gauss1}), lead to
\begin{equation}
\sum_{\mathbf{x}}e^{-i\mathbf{\left(k-k_0\right)}\cdot\mathbf{x}} \exp\left(-%
\frac{\mathbf{x}\cdot\mathbf{x}}{2\sigma^{2}}\right)=  \notag
\end{equation}
\begin{equation}
\sum_{\mathbf{x}}\int_{-\infty}^{\infty}\ldots \int_{-\infty}^{\infty}e^{-i%
\mathbf{\left(k-k_0\right)}\cdot\mathbf{y}} \exp\left(-\frac{\mathbf{y}\cdot%
\mathbf{y}}{2\sigma^{2}}\right)e^{-i2\pi\mathbf{x\cdot y}}\mathbf{dy}=
\notag
\end{equation}
\begin{equation}
\sum_{\mathbf{x}}\int_{-\infty}^{\infty}\ldots \int_{-\infty}^{\infty}e^{-i%
\mathbf{\left(k-k_0+2\pi\mathbf{x}\right)}\cdot\mathbf{y}} \exp\left(-\frac{%
\mathbf{y}\cdot\mathbf{y}}{2\sigma^{2}}\right)\mathbf{dy} .  \label{dos}
\end{equation}
The last integrals can be evaluated using
\begin{equation}
\int_{-\infty}^{\infty} e^{-p^{2}x^{2}\pm q x}{dx}=\frac{\sqrt{\pi}}{p}%
\exp\left(-\frac{q^{2}}{2p^{2}}\right),  \label{tres}
\end{equation}
where $p\geq 0$. In this way we obtain
\begin{equation}
\sum_{\mathbf{x}}e^{-i\mathbf{\left(k-k_0\right)}\cdot\mathbf{x}} \exp\left(-%
\frac{\mathbf{x}\cdot\mathbf{x}}{2\sigma^{2}}\right)=  \notag
\end{equation}
\begin{equation}
\left(\sqrt{2\pi}\sigma\right)^{N}\sum_{\mathbf{x}}e^{-\sigma^{2}\left(%
\mathbf{k-k_0-2\pi\mathbf{x}}\right)^{2}}.  \label{cuatro2}
\end{equation}


\begin{thebibliography}{99}

\bibitem{Aharonov} Y. Aharonov, L. Davidovich, and N. Zagury, \emph{Phys.
Rev. A} \textbf{48}, 1687 (1993).

\bibitem{AAKV01}
D.~Aharonov, A.~Ambainis, J.~Kempe, and U.~Vazirani.
\newblock Quantum walks on graphs.
\newblock In {\em Proc. 33th STOC}, pages 50--59, New York, NY, 2001.

\bibitem{MBSS02}
T.~D. Mackay, S.~D. Bartlett, L.~T. Stephenson, and B.~C. Sanders.
\newblock Quantum walks in higher dimensions.
\newblock {\em Journal of Physics A: Mathematical and General},
\textbf{35}, 2745 (2002).

\bibitem{Tregenna1} B. Tregenna, W. Flanagan, R. Maile, and V. Kendon, \emph{%
New J. Phys.} \textbf{5} 83 (2003).

\bibitem{OPD06}
A.C. Oliveira, R.~Portugal, and R.~Donangelo.
\newblock Decoherence in two-dimensional quantum walks.
\newblock {\em Phys. Rev. A}, \textbf{74}, 012312 (2006).

\bibitem{Watabe08} K. Watabe, N. Kobayashi, M. Katori, and N. Konno, Limit
distributions of two-dimensional quantum walks, Phys. Rev. A \textbf{77}, 062331 (2008).

\bibitem{AF02}
David Aldous and James A. Fill.
\newblock {\em Reversible Markov Chains and Random Walks on Graphs}.
\newblock Monograph at
\url{http://www.stat.berkeley.edu/$\sim$aldous/RWG/book.html}, 2002.

\bibitem{SKW03}
N.~Shenvi, J.~Kempe, and K.B. Whaley.
\newblock A quantum random walk search algorithm.
\newblock {\em Phys. Rev. A}, \textbf{67}, 052307 (2003).

\bibitem{AKR05}
A. Ambainis, J. Kempe, and A. Rivosh.
\newblock Coins make quantum walks faster.
\newblock In {\em Proceedings of the Sixteenth Annual ACM-SIAM Symposium on
  Discrete Algorithms, SODA}, pages 1099--1108, 2005.

\bibitem{PortugalBook}
Renato Portugal.
\newblock {\em Quantum Walks and Search Algorithms}.
\newblock Quantum Science and Technology. Springer, New York, 2013.

\bibitem{FG98}
E.~Farhi and S.~Gutmann.
\newblock Quantum computation and decision trees.
\newblock {\em Phys. Rev. A}, \textbf{58}, 915 (1998).

\bibitem{PRR05a}
A. Patel, K.~S. Raghunathan, and P. Rungta.
\newblock Quantum random walks do not need a coin toss.
\newblock {\em Phys. Rev. A}, \textbf{71} 032347 (2005).

\bibitem{APN13}
A. Ambainis, R. Portugal, and N. Nahimovs.
\newblock Spatial Search on Grids with Minimum Memory.
\newblock {\em arXiv:1312.0172}, 2013.

\bibitem{alejo2010} A. Romanelli, \emph{Phys. Rev. A} \textbf{81}, 062349 (2010).

\bibitem{alejo2012} A. Romanelli, \emph{Phys. Rev. A} \textbf{85}, 012319 (2012).

\bibitem{Zurek} W. H. Zurek, Phys. Rev. D \textbf{24} 1516 (1981); Phys. Rev. D \textbf{26}, 1862 (1982).

\bibitem{Meyer} D. A. Meyer, e-print quant-ph/9804023.

\bibitem{BO78}
Carl M. Bender and Steven A. Orszag.
\newblock {\em Advanced mathematical methods for scientists and engineers}.
\newblock International series in pure and applied mathematics,
McGraw-Hill, New York, 1978.

\bibitem{german2013} M. Hinarejos, A. P\'erez, Eugenio Rold\'an, A.
Romanelli, G.J. de Valc\'arcel, \emph{New J. Phys.} \textbf{15}, 073041 (2013).

\bibitem{Grimmett} G. Grimmett, S. Janson, P.F. Scudo, \emph{Phys. Rev. E}
\textbf{69}, 026119 (2004).

\bibitem{nayak} A. Nayak and A. Vishwanath, e-print quant-ph/0010117

\bibitem{Malena} M. Hor-Meyll, J. O. de Almeida, G. B. Lemos, P. H. Souto
Ribeiro, S. P. Walborn, \emph{Phys. Rev. Letters} \textbf{112}, 053602
(2014).

\bibitem{Carneiro} I. Carneiro, M. Loo, X. Xu, M. Girerd, V. M. Kendon, and
P. L. Knight, \emph{New J. Phys.} \textbf{7}, 56 (2005).

\bibitem{abal} G. Abal, R. Siri, A. Romanelli, and R. Donangelo,\emph{\
Phys. Rev. A} \textbf{73}, 042302, 069905(E) (2006).

\bibitem{salimi} S. Salimi, R. Yosefjani, \emph{Int. J of Mod. Phys. B},
\textbf{26}, 1250112 (2012).

\bibitem{Annabestani} M. Annabestani, M. R. Abolhasani and, G. Abal, \emph{%
J.Phys. A: Math. Theor.} \textbf{43}, 075301 (2010).

\bibitem{Omar} Y. Omar, N. Paunkovic, L. Sheridan, and S. Bose, \emph{Phys.
Rev. A}, \textbf{74}, 042304 (2006)

\bibitem{Pathak} P. K. Pathak, and G. S. Agarwal, \emph{Phys. Rev. A},
\textbf{75}, 032351 (2007)

\bibitem{Petulante} C. Liu, and N. Petulante, \emph{Phys. Rev. A} \textbf{79}%
, 032312 (2009).

\bibitem{Venegas} S. E. Venegas-Andraca, J.L. Ball, K. Burnett, and S. Bose,
\emph{New J. Phys.}, \textbf{7}, 221 (2005).

\bibitem{Endrejat} J. Endrejat, H. B\"{u}ttner, \emph{J. Phys. A: Math. Gen}%
. \textbf{38}, 9289 (2005).

\bibitem{Ellinas1} A.J. Bracken, D. Ellinas, and I. Tsohantjis, \emph{J.
Phys. A: Math. Gen.} \textbf{37}, L91 (2004).

\bibitem{Ellinas2} D. Ellinas, and A.J. Bracken, \emph{Phys. Rev. A} \textbf{%
78}, 052106 (2008).

\bibitem{Maloyer} O. Maloyer, and V. Kendon, \emph{New J. Phys.}, \textbf{9}%
, 87 (2007).

\bibitem{Kubo} R. Kubo, M. Toda, and N. Hashitsume \emph{Statistical Physics
II, Nonequilibrium Statistical Mechanics}, Springer-Verlag, Berlin
Heidelberg New York Tokyo; ISBN 3 540 11461 0, (1985).

\bibitem{Inui} N. Inui, Y. Konishi, and N. Konno, \emph{Phys. Rev. A} \emph{%
Phys. Rev. E}, 052323 (2004).

\bibitem{gustavo} A. Romanelli, G. Segundo,\emph{\emph{Physica A}}, \textbf{393,} 646 (2014).

\end{thebibliography}
\end{document}